# Evidence from Tunneling Spectroscopy for a Quasi-One Dimensional Origin of Superconductivity in $Sr_2RuO_4$

I. A. Firmo[1,2,§], S. Lederer[3,§], C. Lupien[4], A. P. Mackenzie[5,6], J. C. Davis[1,2,5,7] and S.A. Kivelson[3]

1. Laboratory of Solid State Physics, Department of Physics, Cornell University, Ithaca, NY 14853, USA
2. CMPMS Department, Brookhaven National Laboratory, Upton, NY 11973, USA
3. Department of Physics, Stanford University, Stanford, CA 96305, USA.
4. Département de Physique & RQMP, Université de Sherbrooke, Sherbrooke, Québec J1K 2R1, Canada.
5. School of Physics and Astronomy, University of St. Andrews, St. Andrews, Fife KY16 9SS, UK.
6. Max Planck Institute for Chemical Physics of Solids, Nöthnitzer Straße. 40, 01187 Dresden, Germany.
7. Kavli Institute at Cornell for Nanoscale Science, Cornell University, Ithaca, NY 14853, USA.
§   Contributed equally to this project.

**To establish the mechanism of unconventional superconductivity in $Sr_2RuO_4$, a prerequisite is direct information concerning the momentum-space structure of the energy gaps $\Delta_i(k)$, and in particular whether the pairing strength is stronger ("dominant") on the quasi-one-dimensional ($\alpha$ and $\beta$) or on the quasi-two-dimensional ($\gamma$) Fermi surfaces. We present scanning tunneling microscopy measurements of the density-of-states spectra in the superconducting state of $Sr_2RuO_4$ for $0.1T_c < T < T_c$, and analyze them, along with published thermodynamic data, using a simple phenomenological model. We show that our observation of a single superconducting gap scale with maximum value $2\Delta \approx 5\, T_c$ along with a spectral shape indicative of line nodes is consistent, within a weak-coupling model, with magnetically mediated odd-parity superconductivity generated by dominant, near-nodal, Cooper pairing on the $\alpha$ and $\beta$ bands.**

Strong experimental evidence has accumulated that the perovskite superconductor Sr$_2$RuO$_4$ ($T_c$=1.5 K) [1-3] has an unconventional [4], odd parity[5-8] order parameter (OP) that breaks time reversal symmetry [9,10]. A chiral p-wave ("p+ip") state, the quasi-two dimensional analog of the A phase of superfluid $^3$He [11,12], has long been a leading candidate for the order parameter (OP) symmetry of Sr$_2$RuO$_4$. Exotic phenomena, such as topologically protected Majorana edge modes, which are currently the subject of much speculation[13-16], might then be possible in Sr$_2$RuO$_4$. Issues with this OP identification, however, include the apparent absence of the anticipated edge currents [17-20], the absence of a splitting of the transition near $T_c$ by an in-plane magnetic field [21], and the strong evidence [22-30] for lines of gap nodes, or near nodes, that are non-generic in the presence of time-reversal symmetry breaking. We here follow the bulk of the literature and take a chiral p-wave order parameter as a working assumption. (This is discussed further in the Appendix.)

Although, at a microscopic scale, Sr$_2$RuO$_4$ is certainly "strongly correlated" (as evidenced, for instance, by the large mass renormalization [1]) the unconventional superconductivity condenses out of a well-characterized Fermi liquid [1, 31,32] which itself "emerges" from an incoherent metallic regime at a much higher temperature, $T_{FL}$ ~ 30K. This observation suggests [33] that a satisfactory theory of the unconventional superconductivity in this material can be constructed from a weak-coupling perspective. Such a theory would be of great value as a reference point in the ongoing quest to understand unconventional superconductors more generally. Therefore it is important to identify measurements that can, in principle, distinguish between the predictions of different approaches to the superconductivity of Sr$_2$RuO$_4$. One way of doing this is to understand more precisely the structure of the superconducting order parameter, a task made subtle by the multi-band character of Sr$_2$RuO$_4$. In this paper we present scanning tunneling microscope (STM) spectroscopy measurements of the superconducting density-of-states along with theoretical analysis that directly addresses this issue.

The Fermi surface (FS) of Sr$_2$RuO$_4$ (the unit cell is shown in Fig. 1a) consists of three sheets labeled $\alpha$, $\beta$ and $\gamma$ as shown in Fig. 1b [1-3,32]. Hybridization between the two quasi-one-dimensional (1D) bands that originate from the Ru $d_{xz}$ and $d_{yz}$ orbitals leads to the hole-like α sheet and electron-like β sheet, while a band originating from the Ru $d_{xy}$ orbital gives the electron-like, quasi-two-dimensional (2D) $\gamma$ sheet [34-37]. However, the structure of the superconducting energy gaps $\Delta_i(k)$ on these bands, and the identity of elementary interactions generating the pairing, is unknown [1-3]. For instance, in the intermediate-coupling approach of Nomura and Yamada, [38-41] the dominant pairing occurs on the $\gamma$ sheet, while others [42-45], such as the weak-coupling analysis of Raghu *et al.* [46] place it on the α and β sheets. The momentum space locations of the lines of zeros in the gap function, and of the nodes (or near nodes) on the Fermi surface that arise in the latter weak coupling analysis are show in Fig 1b.

I. **STM Results:**

Insight into the gap structures of unconventional superconductors can be obtained from STM-based tunneling spectroscopy, which is particularly suited to the study of materials with low critical temperatures and gap scales below 1 meV [47,48]. In Fig. 2 we show temperature dependent measurements of the differential conductance $dI/dV(V,T) \equiv g(E = eV, T) = \sum_i G_i N_i(E,T)$ on Sr$_2$RuO$_4$ single crystals. Here $N_i(E,T)$ and $G_i$ are, respectively, the tunneling density of states on band $i$ and a weighting factor proportional to the square of the tunneling matrix element between band $i$ and the tip. (Precise definitions are given in the Appendix.) As we will discuss, there are good theoretical reasons to expect $G_\alpha$ and $G_\beta$ to substantially exceed $G_\gamma$.

The samples were grown by the floating zone technique, have $T_c = 1.45$K, and were cleaved in the cryogenic ultra-high vacuum of a dilution-refrigerator based spectroscopic imaging STM (SI-STM) with a lowest temperature of tip electrons ~75mK and the tunneling occurring dominantly along the crystallographic *c* axis. Figure 2a shows a typical topographic image of the RuO$_2$ termination layer of the Sr$_2$RuO$_4$ crystal, showing a disordered surface reconstruction, with short "stripes" of periodicity $4a_0$ running in two orthogonal directions where $a_0 = 3.86$Å is the lattice constant. These surface conditions notwithstanding, the $g(E,T)$ curves measured are found to be independent of location over large fields of view and are therefore apparently uninfluenced by topographic features. Figure 2b shows the dependence on temperature of the g*(E,T)* obtained by averaging over 100 distinct spectra measured at independent locations at each temperature. These spectra can confidently be attributed to the Bogoliubov quasiparticle spectrum of the superconductor for the following reasons: (i) the energy gap observed in $g(E,T)$ disappears at $T_c$ and, concomitantly, g*(0,T)* fills in from 40% at low temperature to 100% of the normal value at $T_c$; (ii) at *T*<100mK, the gap observed in $g(E,T)$ disappears in the presence of a c-axis directed magnetic field of magnitude $\mu_0 H = 100$mT, close to the superconducting $H_{c2}$ ; (iii) for fields $H < H_{c2}$ at $T \ll T_c$, $\Phi$=h/2e vortex cores containing zero-energy states are observed (Fig. 2c).

The data shown in Fig. 2b show several striking features which we analyze more fully later: a) the gap has the pronounced V-shape that indicates the presence of nodes or near-nodes[49]; b) there is no sign of multi-gap structure in the data; c) at the lowest temperatures, the approximate magnitude of the gap maximum $\Delta_{max} \sim 350\ \mu eV$ is comparable to that expected from $T_c$=1.45K within mean field theory. Specific heat and thermal conductivity data discussed below corroborate the existence of nodes or near-nodes and also place a rough lower bound on the size of any subdominant gap.

There are several reasons to expect the $\alpha$ and $\beta$ bands to dominate the c-axis tunneling conductance (*i.e.* $G_\alpha, G_\beta \gg G_\gamma$): In contrast to the $d_{xy}$ orbital, the $d_{xz}$ and $d_{yz}$ orbitals have wavefunction maxima at non-zero values of z, and thus presumably have substantially larger overlap with the tunneling tip. This fact is also reflected in the band structure, as determined by LDA and quantum oscillations: while the $\gamma$ sheet is almost perfectly cylindrical, the $\alpha$ and $\beta$ Fermi surface sheets have greater warping along the c* axis, reflecting much larger interlayer hopping [32]. Although the tip-sample tunneling matrix elements are not the same as those for interlayer hopping, the two parameters reflect similar overlap integrals. Finally, in the related (bilayer) material $Sr_3Ru_2O_7$, for which high-quality surfaces exist with which to identify the contributing bands from quasiparticle interference, the tunneling is clearly dominated by the $d_{xz}$ and $d_{yz}$ orbitals [50]. Altogether, these observations indicate a strong likelihood that the gap shown in Fig. 2b is that on the $\alpha$ and $\beta$ sheets.

## II. Constraints from Bulk Measurements:

Specific heat [23,28] and thermal conductivity [25,26,51] measurements have established the existence of nodes or deep minima in the superconducting gap (from the linear dependence of $C/T$ and the nonzero extrapolation of $\kappa/T$ to zero temperature, respectively). In addition, the fact that $C/T$ extrapolates to nearly zero at zero temperature implies that there is no residual Fermi surface at the lowest temperatures, i.e. that all three bands must host a gap whose magnitude is a substantial fraction of $T_C$. Neither of these qualitative conclusions relies on a mean field picture of superconductivity. If we assume that mean field theory provides a suitable description of the critical point, then we may extract the ratio, $\Delta_{SD}/\Delta_D$, of the subdominant to the dominant gap maxima from the magnitude of the critical jump in the specific heat:

$$\frac{\Delta C}{C} = \frac{\Delta C_D}{C_D}\left[\frac{1 + \eta\frac{\rho_{SD}}{\rho_D}\left(\frac{\Delta_{SD}}{\Delta_D}\right)^2}{1 + \frac{\rho_{SD}}{\rho_D}}\right] \quad (1)$$

Here $\Delta C/C$ is the fractional jump in the specific heat of the full system, $\Delta C_D/C_D$ is the mean field jump of the specific heat contribution by the dominant band(s), $\rho_{SD}/\rho_D$ is the normal state ratio of density states of subdominant band(s) to dominant band(s), and $\eta$ is a dimensionless number (which is typically close to 1) that depends (weakly) on the form factors of the various gaps, suitably averaged on their respective Fermi surfaces. (See the Appendix for full details.)

Low temperature thermal conductivity is another particularly useful probe of the superconducting order parameter. For weak impurity scattering at zero temperature, a line node is expected to give a "universal" contribution to the in-plane thermal conductivity determined only by the quasiparticle dispersion at the nodal point [52]. The observation [51] that the residual thermal conductivity depends only weakly on the residual resistivity suggests the validity of such a picture. However, nodes are experimentally distinguishable from deep near-nodes only if the minimum gap, $\Delta_{min}$, exceeds the energy scales set by the larger of the base temperature $T_0$ and the effective quasiparticle scattering rate $\nu$ at this temperature. The minimum temperature probed by Suzuki et al. is $100\ mK \sim 10\ \mu eV$.

The scattering rate in the superconducting state is difficult to measure, but it is presumably bounded above by the scattering rate in the normal state, as inferred from transport. If we take band structure parameters from quantum oscillations and assume that all three bands have the same scattering rate, we infer a normal state scattering rate of approximately $\nu = 30\ \mu eV$ from resistivity measurements [1] and the Drude formula. Given the observation of universal nodal heat transport, this provides an upper bound on the gap minimum, corresponding to roughly a tenfold anisotropy, $D_{min}/D_{max} < 1/10$.

### III. Phenomenological Model:

The weak coupling analysis of Ref. [46] is consistent with most of the key qualitative features outlined above. The dominant gap is predicted to lie on the α and β sheets, and to contain near nodes [53]. Line nodes in 3D (point nodes in 2D) are generically stable only if time reversal symmetry (or a closely related symmetry) is preserved. The existence of near-nodes in theoretical treatments of p+ip pairing on the quasi-1D bands thus requires explanation.

Even ignoring mixing with the $\gamma$ band, one must still treat the $\alpha$ and $\beta$ bands in a two-orbital basis, where the corresponding Wannier functions correspond to the Ru $d_{xz}$ and $d_{yz}$ orbitals. Thus, the gap function is expressed as a 2 x 2 matrix in the Wannier basis

$$\underline{\Delta}(\vec{k}) = \frac{(1+\tau_3)}{2}\Delta_{xz}(\vec{k}) + \frac{(1-\tau_3)}{2}\Delta_{yz}(\vec{k}) + \cdots \qquad (2)$$

where $\tau_3$ is the diagonal Pauli matrix and … represents small orbital off-diagonal terms which are negligible. The symmetries of a p±ip state imply that

$$\Delta_{xz}(k_x, k_y) = \mp i\Delta_{yz}(k_y, k_x) = -\Delta_{xz}(-k_x, k_y) = \Delta_{xz}(k_x, -k_y) \qquad (3)$$

By appropriate choice of phase, $\Delta_{xz}(\mathbf{k})$ can be taken to be pure real. As the band structure mixes the two orbitals, precise nodes will only occur in fine-tuned cases, but on segments of the Fermi surface which have close to a unique orbital content (*i.e.* those portions of the $\alpha$ and $\beta$ bands far from their avoided crossing in Fig. 1), near nodes can occur with a gap minimum which is parametrically smaller than the typical gap.

Even so, the occurrence of nodes in the absence of orbital mixing is not *required* by symmetry, even if they are not forbidden. Thus, the occurrence of such "accidental nodes," which are well known from studies of unconventional pairing in quasi 1D systems, is a direct reflection of the mechanism of pairing. As an illustrative example, consider a single quasi-1D band (*i.e.* as would arise from the $d_{xz}$ orbital alone). Optimal nesting, and correspondingly a peak in the antiferromagnetic fluctuation spectrum, occurs at momentum $\mathbf{Q}_{1D} = (2k_F, \pi)$. The resulting gap structure which best takes advantage of such interactions satisfies $sign[\Delta_{xz}(k_F, k_y)] = -sign[\Delta_{xz}(k_F, k_y + \pi)]$, thus implying the existence of accidental nodes at $k_{node} \approx (\pm k_F, \pm \pi/2)$ [46,54]. (In $Sr_2RuO_4$, the susceptibility [55] is large at $\mathbf{Q}_{1D}$, but is still larger at $\mathbf{Q}_{max}=(2k_F,2k_F)$ – this affects the precise position of the nodes [33,40], but does not change the reason for their existence.)

For purposes of the explicit calculations performed below we will take the simplest single-harmonic form of the gap function consistent with these general considerations,

$$\Delta_{xz}(k) = \Delta_{1D}[\sin(k_x)\cos(k_y)] \qquad (4)$$

The tight-binding representation of the band structure we adopt includes weak orbital mixing produced by a second neighbor hopping matrix element, $t''$, that transforms the gap nodes into gap minima with gap magnitude of order $\Delta_{1D}(t''/t)^2$, where $t$ is the nearest-neighbor hopping matrix element. For the LDA band structure of Sr$_2$RuO$_4$ [35], these gap minima are on the order of $20\ \mu eV$.

In contrast, the breaking of time-reversal symmetry forbids the existence of nodes on the quasi-2D $\gamma$ band, even in the absence of orbital mixing. Symmetry requires a p-wave gap to vanish at $(\pi, 0)$ and $(0, \pi)$, and most models [38-41,56,57] of chiral p-wave pairing on the $\gamma$ band do indeed have modest gap minima at the closest approach of the Fermi surface to these points. However, these minima are not generically deep enough to account for the observed nodal behavior at low temperatures unless the $\gamma$ band Fermi surface is fine tuned very close to the aforementioned Van Hove points [58]. Moreover, its nearly circular Fermi surface suggests that near nodes are unlikely to occur in any generic circumstances.

The key implication of the above analysis is that the superconducting density-of-states data support a scenario in which superconductivity in Sr$_2$RuO$_4$ originates on the quasi-1D $\alpha$ and $\beta$ bands, with a gap on the quasi-2D $\gamma$ band induced by the proximity effect. To check its consistency with experiment, we now present a simplified phenomenological model with these characteristics and calculate the density of states, electronic specific heat, and low temperature in-plane thermal conductivity. We work in a Wannier basis with annihilation operators $c_x, c_y, c_z$ corresponding to the d$_{xz}$, d$_{yz}$, and d$_{xy}$ orbitals and ignore spin-orbit coupling (and therefore suppress spin indices). The mean-field Hamiltonian for the quasi-1D bands is

$$H_{1D} = \sum_k [\epsilon_{xz}(k) - \mu]c^\dagger_{x,k}c_{x,k} + [\epsilon_{yz}(k) - \mu]c^\dagger_{y,k}c_{y,k} + u_{xy}(k)(c^\dagger_{x,k}c_{y,k} + c^\dagger_{y,k}c_{x,k}) +$$
$$[\Delta_{xz}(k)c^\dagger_{x,k}c^\dagger_{x,-k} + \Delta_{yz}(k)c^\dagger_{y,k}c^\dagger_{y,-k} + \text{h.c.}] \quad (5)$$

with

$$\epsilon_{xz}(k) = -2t\cos(k_x) - 2t^\perp \cos(k_y) \quad \epsilon_{yz}(k) = -2t^\perp \cos(k_x) - 2t\cos(k_y) \text{ and}$$
$$u_{xy}(k) = -4t''\sin(k_x)\sin(k_y)$$

with $\Delta_{xz}$ and $\Delta_{yz}$ from Eqs. (2) and (3), above. When computing the specific heat we take the Hamiltonian for the quasi-2D band to be

$$H_{2D} = \sum_k [\epsilon_{xy}(k) - \mu_z]c^\dagger_{z,k}c_{z,k} + [\Delta_z(k)c^\dagger_{z,k}c^\dagger_{z,-k} + \text{h.c.}] \quad (6)$$

with $\epsilon_{xy}(k) = -2t_z \cos(k_x) - 2t_z \cos(k_y) - 4\tilde{t}_z \cos(k_x)\cos(k_y)$ and

$\Delta_{xy}(k) = \Delta_{2D}[\sin(k_x) + i\sin(k_y)]$.

Unless otherwise specified, the band structure parameters for what follows are taken from LDA [35] and given by $(t, t^\perp, t'', \mu, t_z, \tilde{t}_z, \mu_z) = (1, 0.1, 0.1, 1, 0.55, 0.2, 0.7)$, with the energy scale set by $t = 0.1\ eV$. The a and b bands are "quasi-1D" to the extent that $t^\perp$, the intra-orbital nearest neighbor hopping perpendicular to the orbital axis, is small compared to t. $t''$, which determines the degree of orbital mixing, is the inter-orbital next-nearest neighbor hopping. Note that the assumed form of $\Delta_{xy}$ is the lowest harmonic consistent with p+ip symmetry; the choice is motivated by simplicity rather than any detailed physics.

In analyzing the c-axis tunneling conductance data we assume that tip-sample tunneling occurs only into the quasi-1D bands. We compute the temperature dependent gap amplitude $\Delta_{1D}(T)$ from the BCS gap equation in the simple case of zero orbital mixing using the observed value of $T_c = 1.45$ K to set the overall scale, and include as the single parameter obtained from comparison with the STM tunneling density-of-states data an energy-independent scattering rate $\nu = 70\mu eV$, chosen to yield the correct zero-energy DOS at base temperature[59]. The results are shown in Fig. 3a, with more details on the calculation and more detailed fits described in the Appendix. Larger discrepancies between experiment and the phenomenological model occur at temperatures comparable to $T_C$ where the temperature dependence of the scattering rate and the effects of critical fluctuations are likely significant.

To compare with the specific heat data (Fig. 3b), we take the induced gap on the $\gamma$ band to have a maximum 0.7 times that on the $\alpha$ and $\beta$ bands to fit the critical jump. We take a bulk scattering rate (which controls the low $T$ asymptotic value of $C/T$) to be ½ as large as that at the surface, i.e. 35 $\mu eV$, comparable with the estimate of the normal state scattering rate inferred from transport.

We calculate the low temperature in-plane thermal conductivity, $\kappa/T$, using the formula of Durst and Lee [52], adapted to our scenario in which the $\beta$ band has eight gap minima with $D_{min} < n$ (which are thus experimentally indistinguishable from nodes).

Taking the Fermi velocity of the $\beta$ band from quantum oscillations [1] and the gap magnitude from mean field theory as above, we obtain a predicted value of k/T= 12.5 $mW(K^2 cm)^{-1}$, comparable to but smaller than the experimental value of 17 $\pm$ 1.5 $mW(K^2 cm)^{-1}$[51]. The discrepancy is not alarming, given the crudeness of the model adopted and the asymptotic character of the Durst-Lee analysis. For instance, for the bulk scattering rate and gap magnitudes obtained from our fits to the specific heat, there are an additional 8 locations in the Brillouin zone (in addition to the minima on the $\beta$ band) where the gap is less than twice $\nu$: on the $\gamma$ band along the (1,0) and (0,1) directions, and on the $\beta$ band along the (1,1) and (1,-1) directions. Contributions from the neighborhood of these points could account for an enhancement of k/T. If this is the case, experiments performed on samples of increasing purity will exhibit a reduced asymptotic value of k/T once the scattering rate is significantly smaller than the secondary gap extrema, and in any case will vanish once the scattering rate (and the base temperature) drop below $D_{min}$ (which is on the order of 20 $\mu eV$ in this model).

The good qualitative agreement with both experiments shown in Figs. 3 confirms the plausibility of the proposed phenomenological picture. We also note that the values of the scattering rates required for the fits are small and realistic, i.e. of a magnitude consistent with $T_c$ = 1.45 K in this extremely purity-sensitive superconductor [4].

IV. **Summary:**

We have presented qualitative evidence of dominant, accidentally near-nodal superconductivity on the quasi-1D bands, and have shown, using a simple phenomenological model, the consistency of this picture with thermodynamic and spectroscopic evidence. Direct confirmation that the superconducting density-of-states in STM tunneling is dominated by the quasi 1D bands – the only speculative step in the analysis—could be obtained from an observation of the predicted pattern of Bogoliubov quasiparticle interference (QPI). This technique, which has yielded detailed information about the superconducting gap structure of the cuprates [60], iron pnictides [61] and heavy fermion superconductors [47], has not yet been applied to $Sr_2RuO_4$. Our findings provide immediate motivation for such a study.

*Note added in proof:* A new weak coupling calculation which treats all three bands and the spin-orbit coupling between them non-perturbatively has recently been carried out by Scaffidi *et al.*; among other things, they find a regime in which the gaps on all three bands are comparable in magnitude, while on the $\alpha$ and $\beta$ bands, the gap function has deep accidental near-nodes, consistent with the earlier calculaitons of Raghu *et al.*[46] and what has been assumed in the present paper.

**Acknowledgements** We especially acknowledge and thank Y. Maeno for access to his high quality $Sr_2RuO_4$ crystals and for many constructive comments. We thank M. Aprili, S.-B. Chung, A. Damascelli, F. Mark Fischer, E.-A. Kim, I. Mazin, S. Raghu, Jean-Philippe Reid, T.M. Rice, and A.M. Rost for helpful discussions and communications. Studies at Cornell University were supported by the Office of Naval Research under Award N00014-13-1-0047, and at Stanford under DE-AC02-76SF00515. APM acknowledges the receipt of a Royal Society-Wolfson Research Merit Award and the support of EPSRC through the Programme Grant 'Topological Protection and Non-Equilibrium States in Correlated Electron Systems". IAF acknowledges support from Fundação para a Ciência e a Tecnologia, Portugal under fellowship number SFRH/BD/60952/2009.

**Author Contributions**: C.L., I.A.F & S.L carried out the SI-STM experiments plus the data preparation and analysis; Y. M. synthesized and characterized the samples; A.P.M., J.C.D. and S.A.K supervised the project, and wrote the paper with key contributions from S.L and I.A.F. The manuscript reflects contribution and ideas of all authors.

* To whom correspondence should be addressed: sakivelson@stanford.edu

Figure 1

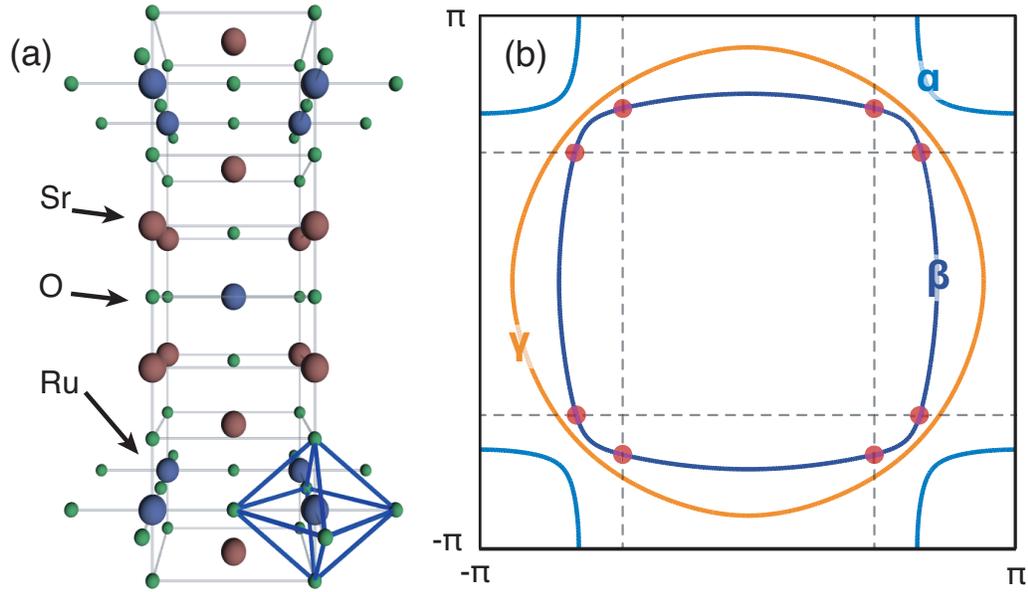

**Figure 1.** Crystal structure and Fermi surface of $Sr_2RuO_4$:

(a) Unit cell of $Sr_2RuO_4$. Sr atoms are shown in red, Ru atoms in blue and O atoms in green.

(b) The $\alpha$- and $\beta$- bands are shown in two shades of blue. The quasi-2D $\gamma$-band is shown in orange. We neglect the dispersion along $k_z$ and therefore present a two-dimensional cross section of the Fermi surface. The approximate locations of the lines of zeros in the gap function of Ref. 39 are shown as dashed lines. The near-nodes expected on the $\beta$ band are then indicated by 8 red dots.

Figure 2

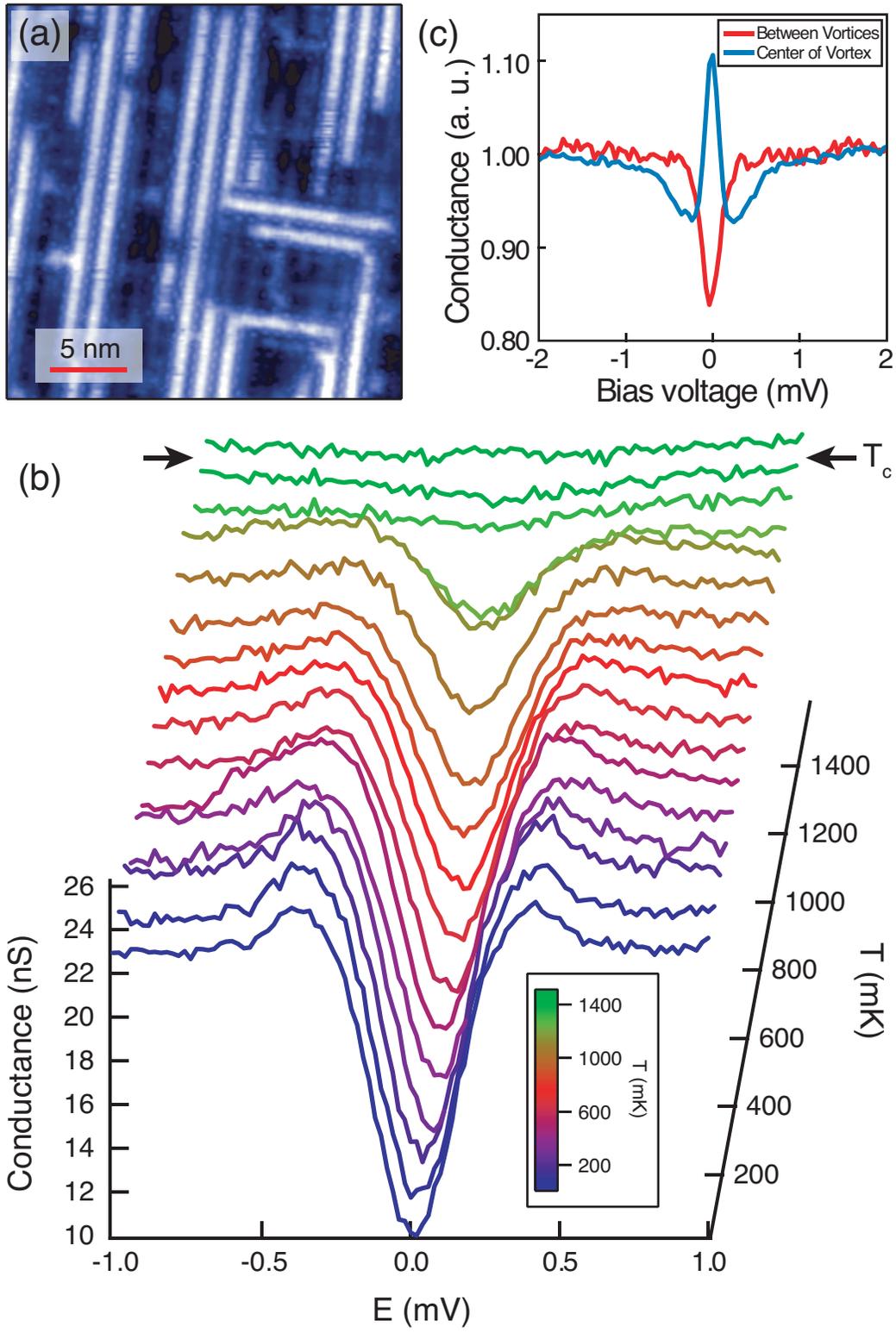

**Figure 2.** Tunneling density-of-states results

(a) Topographic image of the $RuO_2$ plane of $Sr_2RuO_4$ acquired at a 2GΩ junction resistance, -100mV tip-sample bias.

(b) Differential conductance spectra for a sample temperature range between 20 mK and 1.5 K. The minimum electron temperature in the tip is ~75mK. The observed gap becomes zero above the superconducting $T_c$ = 1.45 K, as one would expect for the superconducting gap, $\Delta(T)$. A finite N(E=0) density of states at $E_F$ is observed at all T. The shape of this spectrum is very consistent with a nodal gap structure and the gap magnitude of ~350 meV, consistent with $k_B T_c$, must then be that of the primary gapped Fermi surface. The arrows represent $T_c$. The data at higher temperature were normalized to the same normal state conductance (at E >> $\Delta$) as the 21mK data.

(c) At B=0.15T, the measured N(E) everywhere far from vortex cores is in red and the quasiparticle bound states within the vortex cores in blue (which from their areal density exhibit $\Phi = h/2e$ as shown by imaging the vortex-core locations). The data were acquired at 13 MΩ junction resistances, 2mV tip-sample bias.

Figure 3

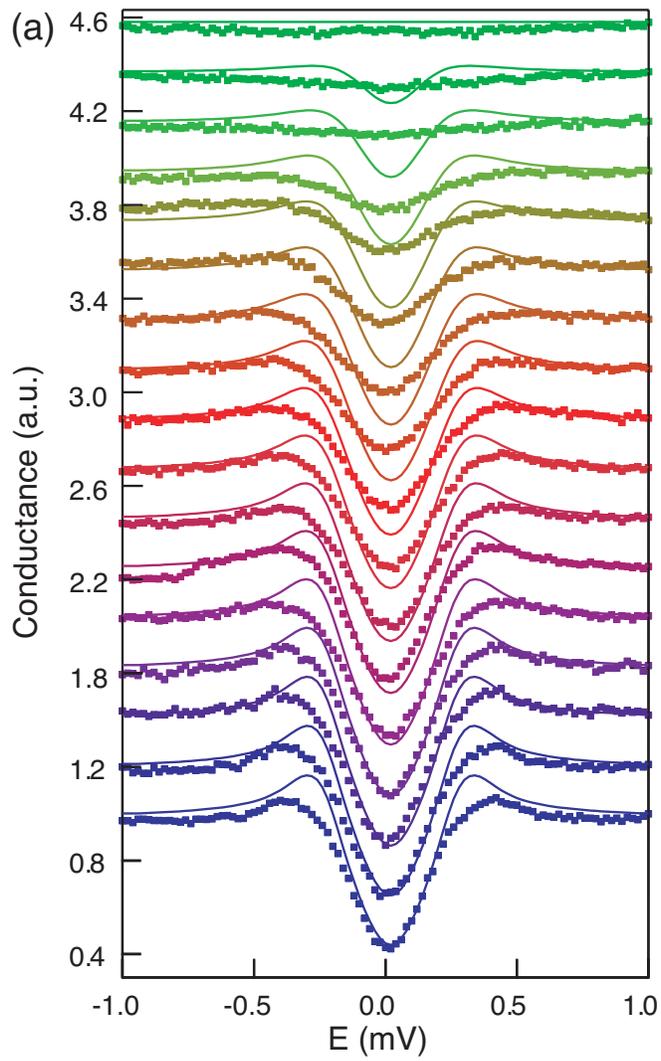

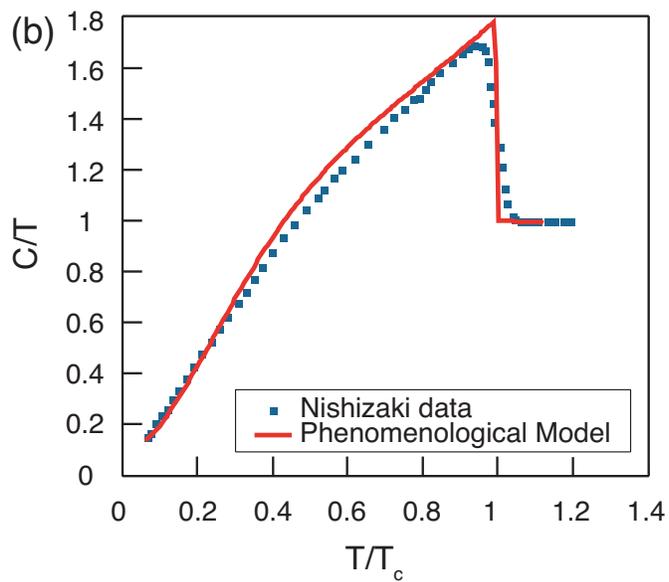

**Figure 3.** Phenomenological model applied to tunneling DOS and specific heat data:

(a) Comparison of normalized tunneling conductance data with the predictions of the phenomenological model. A single parameter, an energy- and temperature-independent scattering rate $\nu = 70\mu\text{eV}$, has been introduced and set to fit the zero-bias conductance at base temperature. The colors indicate the temperature of a given data point, where the same color scale was used as in Fig. 2(b).

(b) Comparison of published specific heat data (Nishizaki et al [23]) with the predictions of the phenomenological model with parameters described in the main text. C/T has been normalized to the normal state value and T to the critical temperature.

**Appendix A. Tunneling density-of-states calculation**

We first consider the case of a single band superconductor described by a retarded Nambu Green's function $\mathbf{G}(\mathbf{k}, \omega)$. A standard calculation gives the "tunneling density of states" as

$$N(eV) \equiv -\int d\omega \, f'(\omega - eV)\rho_e(\omega) \tag{1}$$

where $f(x) \equiv [1 + \exp(-x/T)]^{-1}$ is the Fermi function, and

$$\rho_e(\omega) \equiv -\frac{2}{\pi}\int \frac{d^2k}{(2\pi)^2} Im[G_{11}(\mathbf{k}, \omega)] \tag{2}$$

(the 2 is for spin). The function $\rho_e$ is not equal to the density of Bogoliubov quasiparticle states defined as

$$\rho_{qp}(\omega) \equiv -\frac{1}{\pi}\int \frac{d^2k}{(2\pi)^2} Im[\mathbf{G}_{11}(\mathbf{k}, \omega) + \mathbf{G}_{22}(\mathbf{k}, \omega)], \tag{3}$$

which in the absence of disorder is related to the quasiparticle dispersion, $E_k$:

$$\rho_{qp}(\omega) = \frac{1}{2}\int \frac{d^2k}{(2\pi)^2}[\delta(\omega - E_k) + \delta(\omega + E_k)] \tag{4}$$

However, the difference between $\rho_e$ and $\rho_{qp}$ is negligible so long as the Fermi energy is large compared to $\Delta$ and $\omega$. Since the range of bias energies shown in the paper and the low-temperature gap observed in $Sr_2RuO_4$ are both on the order of $10^{-3}t$, we will use the simpler expression $\rho_{qp}$ in what follows. Thus, for present purposes, we approximate $\rho_e$ by $\rho_{qp}$ in Eq. (1).

To account for the effects of disorder in the simplest phenomenological manner, we include an elastic scattering rate $\nu$ in the Green's function by taking $\omega \to \omega + i\nu$. This broadens the delta functions to Lorentzians:

$$\rho_{qp}(\omega) = \frac{1}{(2\pi)^3} \int d^2k \frac{\nu}{(\omega - E_k)^2 + \nu^2} + \frac{\nu}{(\omega + E_k)^2 + \nu^2} \qquad (5)$$

The extension to the multi-band case is now straightforward, and involves adding terms in the integral above corresponding to additional branches of the quasiparticle dispersion. For the parameters used, the effect of orbital mixing on the tunneling density of states is negligible, so the curves shown are the result of a single band calculation corresponding to only the $d_{xz}$ orbital. In addition, the final curves are normalized to their values at $1\ m$eV (i.e. the normal state value).

**Appendix B. Alternate DOS Fits**

In the body of the paper we present calculations with a single free parameter, a temperature- and energy-independent scattering rate $\nu_0$ chosen to reproduce the zero-bias DOS at base temperature. We now present three more sets of theoretical curves which incorporate additional, physically reasonable parameters: a quadratic temperature dependence of the scattering rate and a rescaling of the gap $\Delta_{1D}(T)$. In Figure 1a) we have the curves as shown in the paper, with the gap given by mean field theory with a scattering rate $\nu_0 = 70\mu$eV. In Figure 1b) we again take the gap given by mean field theory, but now take $\nu(T) = \nu_0 + AT^2$, with $A = 10\ \mu eV/K^2$. In Figure 1c) we take the conditions of Fig. 1a, but rescale both the gap and the scattering rate by a temperature independent factor of 1.14. In Figure 1d), we take the conditions of Fig. 1b, but rescale both the gap and the scattering rate by a temperature independent factor of 1.14.

**Appendix C. Specific Heat Calculations**

The expression for the electronic specific heat of a single-band superconductor in the clean limit is well known:

$$C = \frac{-2}{T} \int \frac{d^2k}{(2\pi)^2} f'(E_k) \left[ E_k^2 - \frac{1}{2} T \frac{\partial |\Delta_k|^2}{\partial T} \right] \tag{6}$$

$$= \frac{-2}{T} \int \frac{d^2k}{(2\pi)^2} \int d\omega \, [\delta(\omega - E_k)] f'(\omega) \left[ \omega^2 - \frac{1}{2} T \frac{\partial |\Delta_k|^2}{\partial T} \right]$$

Suppose the momentum and temperature dependence of the gap are given by $|\Delta_k|^2 = |\Delta(T)|^2 g_k^2$, where $g_k$ is a dimensionless form factor whose maximum value on the Fermi surface is 1. Then the fractional jump in the specific heat at the critical temperature is

$$\frac{\Delta C}{C} = \frac{-\frac{T_C}{2} \frac{d|\Delta|^2}{dT} \int d^2k \, f'(|\epsilon_k|) g_k^2}{\int d^2k \, f'(|\epsilon_k|) \epsilon_k^2} = -\frac{3}{2 \pi^2 T_C} \frac{d|\Delta|^2}{dT} \langle g_k^2 \rangle$$

where the second equality is in the weak coupling limit (so that we can linearize the dispersion about the Fermi surface), and the angle brackets represent a Fermi surface average:

$$\langle X(k) \rangle := \frac{\oint \frac{X(k)}{|v_k|} dk}{\oint \frac{1}{|v_k|} dk}$$

If the gap is an intrinsic or "dominant" gap $\Delta = \Delta_D$ (i.e. not induced by the proximity effect), then the temperature dependence of the gap magnitude is determined by the BCS gap equation, which yields

$$\frac{1}{T_C} \frac{d|\Delta_D|^2}{dT} = -a \frac{\langle g_k^2 \rangle_D}{\langle g_k^4 \rangle_D}$$

where the subscript $D$ refers to the gap and the Fermi surface of the dominant band and

$$a = 8\left(\int_0^\infty dx \left[\frac{\tanh x}{x^3} - \frac{\text{sech}^2 x}{x^2}\right]\right)^{-1} = 9.38$$

The fractional specific heat jump of the dominant band is therefore

$$\frac{\Delta C_D}{C_D} = -\frac{3}{2\pi^2}\left(\frac{1}{T_C}\frac{d|\Delta_D|^2}{dT}\right)\langle g_k^2\rangle_D = \frac{3a}{2\pi^2}\frac{\langle g_k^2\rangle_D^2}{\langle g_k^4\rangle_D} = 1.43\frac{\langle g_k^2\rangle_D^2}{\langle g_k^4\rangle_D}$$

Now consider a "subdominant" gap with a different form factor $h_k$, whose maximum $\Delta_{SD}(T)$ is a temperature-independent fraction $r$ of that of the dominant gap $\Delta_D$. Then we can easily solve for the critical jump in its specific heat at $T_C$

$$\frac{\Delta C_{SD}}{C_{SD}} = -\frac{3}{2\pi^2}\left(\frac{1}{T_C}\frac{d|\Delta_{SD}|^2}{dT}\right)\langle h_k^2\rangle_{SD} = 1.43\left|\frac{\Delta_{SD}}{\Delta_D}\right|^2 \frac{\langle g_k^2\rangle_D \langle h_k^2\rangle_{SD}}{\langle g_k^4\rangle_D}$$

The fractional specific heat of the whole system is then

$$\frac{\Delta C}{C} = \frac{\Delta C_D}{C_D}\left[\frac{1 + \eta\frac{\rho_{SD}}{\rho_D}\left|\frac{\Delta_{SD}}{\Delta_D}\right|^2}{1 + \frac{\rho_{SD}}{\rho_D}}\right]$$

Where we have introduced the ratio of normal state density of states of the subdominant band to that of the dominant band $\frac{\rho_{SD}}{\rho_D}$, and

$$\eta := \frac{\langle h_k^2\rangle_{SD}}{\langle g_k^2\rangle_D}$$

From the data of Nishizaki et al, we estimate $\frac{\Delta C}{C} = 0.75 \pm 0.05$. For the form factors and band structure considered, this yields $\frac{\Delta_{SD}}{\Delta_D} = 0.76 \pm 0.06$.

We account for the effects of disorder in the simplest phenomenological fashion by including a Lorenzian broadening:

$$C = -\frac{2}{T}\int \frac{d^2k}{(2\pi)^2}\int \frac{d\omega}{\pi}\left[\frac{\nu}{(\omega-E_k)^2+\nu^2}\right]f'(\omega)\left[\omega^2 - \frac{T}{2}\frac{d|\Delta_k|^2}{dT}\right] \quad (7)$$

As with the density of states, orbital mixing between the $d_{xz}$ and $d_{yz}$ orbitals has a negligible effect on the specific heat, so single band models were used to compute the contributions of both the quasi-1D and quasi-2D bands. Each specific heat contribution is normalized to its value at $T_C$, and the final result is a weighted average of the quasi-1D and quasi-2D contributions, with the weights (43% and 57% respectively) taken from quantum oscillations[1].

**Appendix D. Predictions for Bogoliubov Quasiparticle Interference (QPI)**

Figure 2 shows curves of constant quasiparticle energy on the $\alpha$ and $\beta$ bands for several values near zero energy. Orbital mixing is included in this calculation, as it shifts the nodes and lifts them to parametrically small gap minima. The near nodes are clearly visible near the zone diagonals for the lowest energy curves, while at energies greater than the gap scale the constant energy contours encompass the Fermi surface of the $\alpha$ and $\beta$ bands.

**Appendix E. Concerning the assumed p+ip symmetry of the pairing**

In the body of this paper, we have focused our analysis on the STM and specific heat data, and to a lesser extent on the thermal conductivity. There is, of course, a large body of other measurements which shed light on the nature of the pairing. In particular, in our analysis, we have *assumed* that the pairing has $p_x+ip_y$ symmetry, and have analyzed these experiments to determine more detailed features of the structure of the gap along the Fermi surface. However, as we have mentioned, while there is considerable direct evidence that this assumption is valid, there are some aspects of certain existing experiments which are not easily reconciled with it. We have implicitly assumed that these discrepancies will be resolved by future theoretical and experimental work in a way that does not overthrow the general consensus concerning the symmetry of the pairing, but as this is not certain, we feel it is worthwhile summarizing the arguments in favor of this conclusion. We relegate this discussion to the Appendix as it is neither strikingly original (see especially the discussions in [63] and, more recently, [3]), nor entirely unassailable.

1) There is strong direct experimental evidence that the superconducting state in $Sr_2RuO_4$ is chiral and time-reversal symmetry breaking – this has been detected optically[10], with $\mu$SR[9], and with various phase sensitive measurements[7,8]. The major worry concerning this conclusion is that the edge currents that are theoretically expected to be observable in such a state have so-far eluded detection[19,20], despite experiments with two orders of magnitude more sensitivity than would be needed to detect currents of the predicted magnitudes. While we do not have any explicit resolution of this contradiction to offer, we consider it likely that the discrepancy reflects a theoretical oversight – for instance, a proper account of the effects of disorder (possibly associated with the surface). If we accept the chiral character of the superconducting state, this implies (except under exotic circumstances) that it derives from a pairing symmetry corresponding to a two-dimensional representation of the point group. For the tetragonal symmetry of $Sr_2RuO_4$, this leaves us the choice of triplet (or more precisely, in the presence of spin-orbit coupling, odd parity) pairing with ($p_x+ip_y$) symmetry, or singlet pairing with ($d_{xz}+id_{yz}$) symmetry.

2) On symmetry grounds, for either form of pairing, the addition of a field which breaks the tetragonal symmetry (e.g. an in-plane magnetic field or uniaxial pressure) the superconducting transition would be expected to either be split into a sequence of two separate continuous transitions, or to be first order. For temperatures not too low compared to the zero field $T_c$, neither of these possibilities arise – in the presence of an applied in-plane magnetic field, there is a field dependent depression of $T_c$, but no splitting of the transition[21]. This is another puzzle (which applies equally to the d and p wave cases) for which we have no compelling explanation – perhaps the splitting is too small to have been observed.

3) Strong NMR evidence[5] (T independence of the Knight shift through $T_c$) indicates spin-triplet pairing – and thus $p_x+ip_y$. If this evidence is set aside, then as has been suggested by [5554] ($d_{xz}+id_{yz}$) pairing remains a viable candidate.

4) The existence of half-quantum vortices as topologically stable excitations is one of the most exciting features of a chiral p+ip superconductor – a feature that arises from its spin rather than from its structure and so is not an expected feature of a d+id superconductor. As has been pointed out by [], energetic considerations imply that isolated half-quantum vortices cannot occur in macroscopic samples, but they can in mesoscopic samples. Thus, the observation [65] in magnetization steps of presumed signatures of half-quantum vortices in mesoscopic samples of $Sr_2RuO_4$ is independent evidence of the p+ip character of the superconducting state.

5) There is evidence from specific heat, thermal conductivity, and other measurements that there are line nodes on the Fermi surface of $Sr_2RuO_4$. Line nodes are typically unstable in a time-reversal symmetry breaking superconductor, although strikingly, the oddness of the d-wave state under reflection through the xy plane implies the existence of ``horizontal'' line-nodes around the cylindrical Fermi surfaces along the lines at which they intersect the $k_z = 0$ and $k_z = \pi$ planes. This is a prima-facie argument in favor of d+id pairing. However, as we have emphasized in the body of this paper, in the case in which the pairing arises from a strongly k-dependent interaction, the multi-band character of the resulting pairing contains parametrically deep ``near nodes'' which under a broad range of circumstances can account for the apparently nodal behavior seen in experiment.

6) As previously mentioned, the most stringent constraints on the gap anisotropy come from studies of the in-plane thermal conductivity[21], since these extend down to quite low temperatures, $T_{base}$=100mK. The fact that k/T approaches a T independent value k/T ~ 1.7W/K$^2$m at low temperature implies the existence of nodes or near nodes with a minimum gap that is smaller than the greater of $\nu$ (the quasiparticle scattering rate) and k$T_{base}$, i.e. $\Delta_{max}/\Delta_{min} > max[\Delta_{max}/\nu, \Delta_{max}/T_{base}]$ Taking the estimate of the gap maximum we have obtained from STM, $\Delta_{max}$ =320meV, yields $\Delta_{max}/T_{base}$= 35. Suzuki *et al* obtained an estimate of the normal state scattering rate, $\nu_{Suzuki}$=7meV, from a mean-field analysis of the superconducting $T_c$; this would imply a very stringent lower bound on the gap anisotropy set by the measurement temperature since $\Delta_{max}/\nu_{Suzuki} > \Delta_{max}/T_{base}$; to make this compatible with an assumed p+ip pairing symmetry would seem to require more gap anisotropy than can be obtained without considerable fine-tuning of parameters. For instance, the ratio of the gap minimum to the gap maximum on the $\beta$ band in the phenomenological model analyzed in the body of this paper yields $\Delta_{max}/\Delta_{min}$ =16. However, the Suzuki *et al* estimate of $\nu$ is indirect. Our analysis of the STM data leads to an estimate of the scattering rate at the surface of $\nu_{STM}$ =70meV or $\Delta_{max}/\nu_{STM}$=4.6 while the specific heat analysis yields an estimate of the scattering rate in the bulk of $\nu_{SpecH}$=35meV or $\Delta_{max}/\nu_{SpecH}$=9. Perhaps the most direct way to estimate the scattering rate is from the normal state resistivity, which for $\rho = 0.1 m\Omega\ cm$ and under the assumption that the scattering rate is the same for all three bands, yields the estimate $\nu_\rho$ =30meV or $\Delta_{max}/\nu_\rho$=11. If we adopt any of these estimates of the scattering rate, then the gap anisotropy in our phenomenological model is compatible with the thermal conductivity data.

7) The highly two-dimensional character of the electronic structure of $Sr_2RuO_4$ provides a strong theoretical argument against any pairing symmetry with a horizontal line of gap nodes. In real-space, what this would imply is that the gap function, $\Delta(\mathbf{r}, \mathbf{r}')$ (or the expectation value of the pair-field operator) vanishes for any two points **r** and **r'** in the same plane. While it is possible to imagine pairing interactions that give rise to this sort of pairing[55], it is rather unnatural from a purely theoretical perspective.

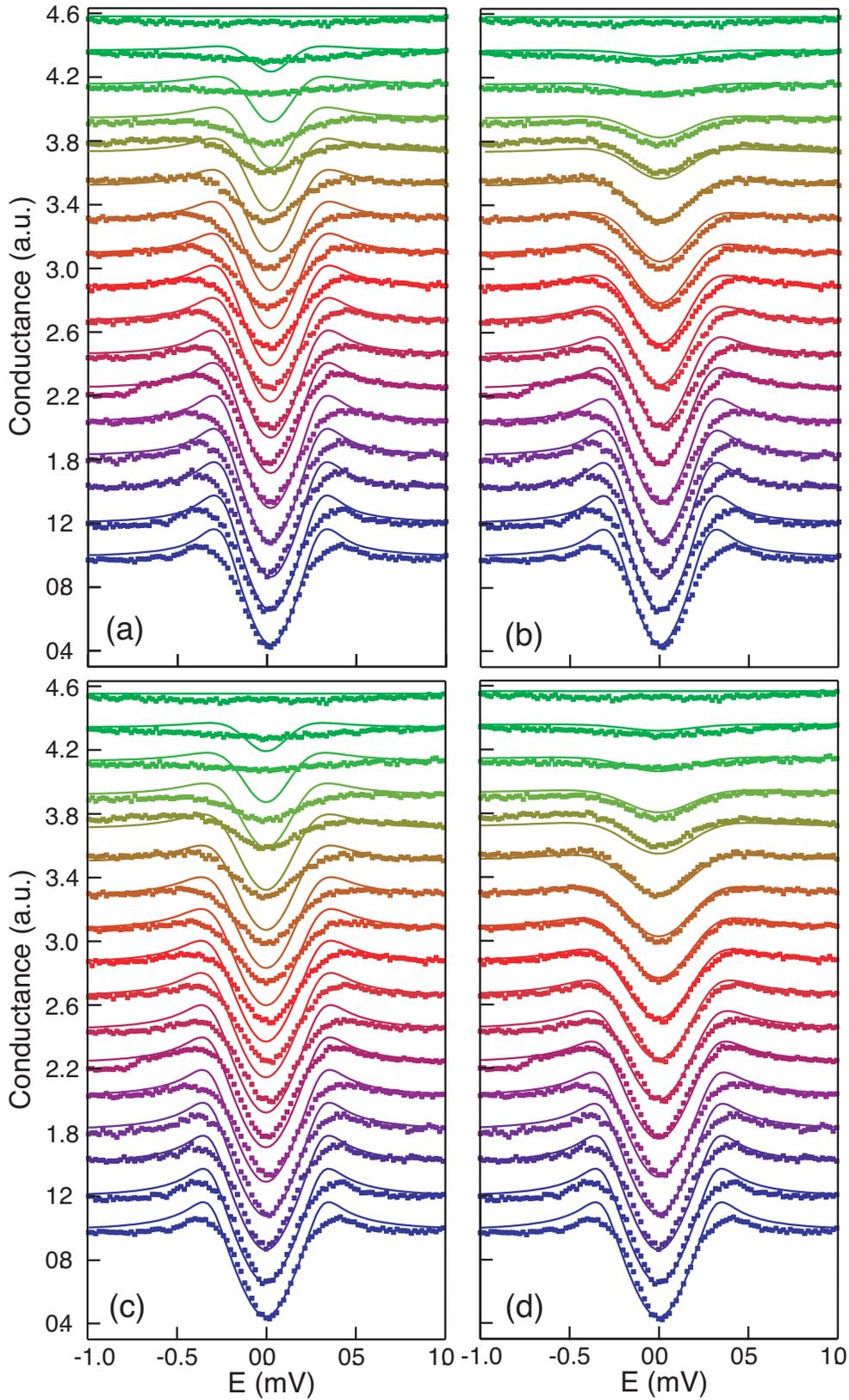

**Figure 1.** Alternate fits to the tunneling data:

(a) Mean field gap with $T_C = 1.45$ K, temperature-independent scattering rate $\nu_0 = 70\mu eV$.

(b) Mean field gap with $T_C = 1.45$ K, temperature-dependent scattering rate $\nu(T) = \nu_0 + AT^2$ with $\nu_0 = 70\mu eV, A = 10\mu eV/K^2$.

(c) The conditions of figure (a) with the gap and scattering rate rescaled by a factor of 1.14.

(d) The conditions of figure (b) with the gap and scattering rate rescaled by a factor of 1.14.

Appendix Figure 2

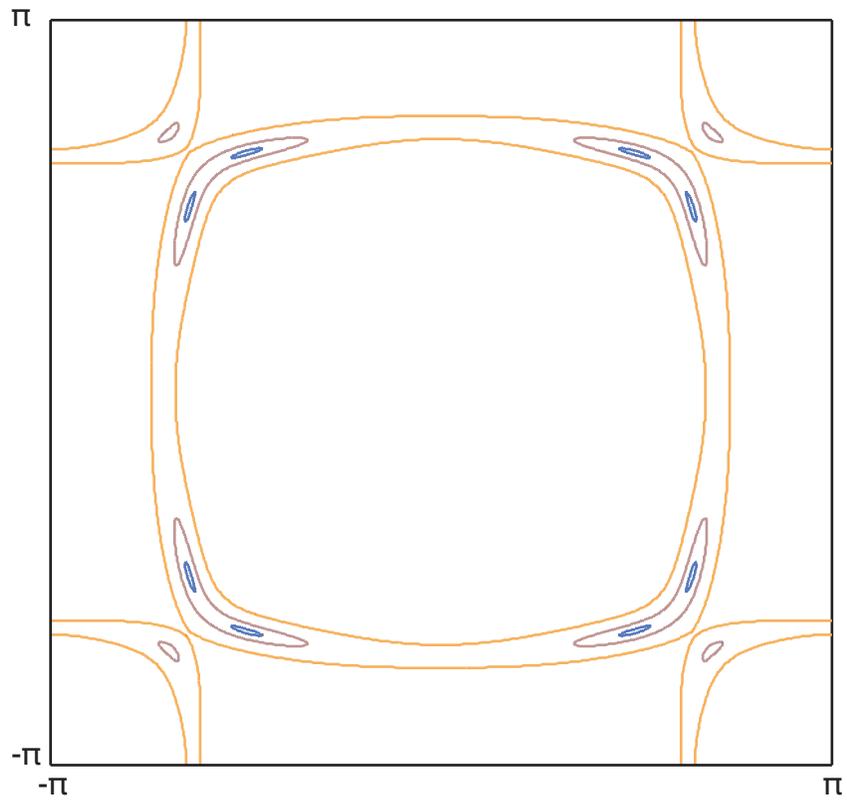

**Figure 2.** Quasiparticle dispersion

Curves of constant quasiparticle energy for several low energy values. Low energy quasiparticle interference should be dominated by scattering between the eight near-nodes situated near the zone diagonals.